# A mesoscale model for quantitative phospholipid monolayer simulations at the air-water interface

Yongzheng Zhu [1, 3, a], Xuan Bai [2, a], Guoqing Hu [2, *]

[1] The State Key Laboratory of Nonlinear Mechanics (LNM), Institute of Mechanics, Chinese Academy of Sciences, Beijing 100190, China

[2] Department of Engineering Mechanics & State Key Laboratory of Fluid Power and Mechatronic Systems, Zhejiang University, Hangzhou 310027, China

[3] School of Engineering Science, University of Chinese Academy of Sciences, Beijing 100049, China

*a Equal contribution.*

**ABSTRACT:**

A mesoscale model with molecular resolutions is presented for the dipalmitoyl-phosphatidylcholine (DPPC) and 1-palmitoyl-2-oleyl-sn-glycero-3-phosphocholine (POPC) monolayer simulations at the air-water interface using many-body dissipative particle dynamics (MDPD). The parameterization scheme is rigorously based on reproducing the physical properties of water and alkane and the interfacial property of the phospholipid monolayer by comparing with our experimental results. The MDPD model yields a similar surface pressure-area isotherm as well as the similar pressure-related morphologies compared with the all-atomistic simulations and experiments. Moreover, the compressibility modulus, order parameter of lipid tails, and thickness of the MDPD phospholipid monolayer are quantitatively in line with the all-atomistic simulations and experiments. This model can also capture the sensitive changes in the pressure-area isotherms of the mixed DPPC/POPC monolayers with altered mixed ratios by comparing with the experiments, indicating that our model scheme is promising in applications for complex natural phospholipid monolayers. These results demonstrate a significant improvement on quantitative phospholipid monolayer simulations over the previous coarse-grained models.

# 1. INTRODUCTION

Lipid monolayers at the air-water interface are of interest in a variety of disciplines. Through the correspondence between the lipid monolayer and bilayer, the interpretation of the lipid bilayer properties can be obtained from the lipid monolayer experiments that are more easily performed than the bilayer experiments (1-4). Besides, studying the properties of the lipid monolayer is crucial of understanding the biophysical function of the lung surfactant monolayer. Lung surfactant is consisted of hundreds of lipids (~90% by weight), mainly are dipalmitoylphosphatidylcholine (DPPC) and unsaturated phosphatidylcholines (PCs), and 4 types of surfactant proteins (~10% by weight) (5-7). It can adsorb on the surface of the alveoli fluids that reduces the surface tension of the alveoli to maintain the tidal respiration (8, 9).

Studying the structure and mechanical properties of the lipid monolayer during compression and expansion is a central question of the biophysics of the lung surfactant (10-12). To date, experimental studies can measure the phase coexistence, compressibility and surface tension of the lipid monolayer through the AFM, Langmuir- Blodgett balance (LB), the captive bubble methods (CB) and other techniques (13-17). However, it is still difficult to directly study and observe the mesoscopic details through these experiments, but these can be obtained from the molecular dynamics (MD) simulations (18-21). Therefore, MD simulations have an immeasurable potential for a deeper understanding of the lipid monolayer at the molecular level.

At present, the commonly used MD methods for studying biomembrane systems including all-atom (AA) MD and coarse-grained (CG) MD. Although the AAMD, such as CHARMM force field (22), has the high calculation accuracy, it takes a lot of computing resources and time to simulate biological systems with a spatial scale of more than tens of nanometers and is thus not suitable for simulating the mesoscopic phenomena of the lipid monolayer. In CGMD simulations, a cluster of atoms is considered as one bead that interacts with each other, therefore decreasing the freedom of the total atoms and saving lots of computing costs. Nowadays, Martini force field (23, 24) CGMD and dissipative particle dynamics (DPD) (25) are two most popular CGMD simulation methods for biological systems. The Martini force field provides systematic force field parameters for commonly used lipids and other biological molecules and thus is suitable for modeling the complex lipid monolayers at the air/water interface. However, the Martini water models evidently underestimate the air/water surface tension due to the narrow well of the Lennard-Jones (LJ) potential, which is limited to modeling interfacial adsorption, pore formation and the pressure-area isotherm of the monolayers. DPD was initially proposed for simulating complex fluids at the mesoscopic scale and has also been widely used for biomembrane simulations (26). Compared with the Martini force field, DPD has a soft potential that allows a larger timestep in simulations and is more suitable for modeling the mesoscopic thermodynamics due to the included dissipative forces and random forces. However, the accuracy and the universality of DPD is much weaker than Martini force field. Most importantly, DPD cannot handle the problems of the air-water interface due to the absence of the attractions in its interaction potential.

The many-body (or multi-body) dissipative particle dynamics (MDPD) modifies the original DPD by replacing the purely repulsive conservative forces by forces deriving from a many-body potential (27). In this way, the equation of state has a higher-order pressure-density curve to accommodate vapor-liquid coexistence than DPD models. Thus, the conservative force of the MDPD was modified to be related to density (28). Moreover, the conservative force of MDPD was further developed by adding a pair of soft attractive interactions with a larger cut-off radius than the repulsive interactions (29). Recently, MDPD has been used for studying the adsorption behavior of simple surfactants (30, 31). However, the MDPD model for quantitative lipid simulations is still relatively complicated.

Thus, we proposed MDPD models for two commonly used phospholipids, that is DPPC and 1-palmitoyl-2-oleyl-sn-glycero-3-phosphocholine (POPC), for lipid monolayer simulations. The parametrization for the lipid models is rigorously based on reproducing the experimental density and surface tension of reference systems (water and alkane) and the surface pressure-area isotherm of the lipid monolayer. We then examine the mechanical property (compressibility modulus) and the molecular structures, including pore formation, collapse, thickness, and order parameter of the modeled lipid monolayers by comparing with the experiments and other simulations. Finally, we test the expansibility and compatibility of our models by simulating the mixed DPPC/POPC monolayer at different ratios, where the pressure isotherm fits well with the experimental results.

## 2. MATERIALS AND METHODS

*2.1. Multi-body dissipative particle dynamics simulations (MDPD)*

### 2.1.1. Theory and algorithm

The motion of the MDPD particles is described according to Newton's second law by the conservative force, dissipative force, and random force.

$$\frac{d\mathbf{v}_i}{dt} = \mathbf{F}_i = \sum_{i \neq j}\left(\mathbf{F}_{ij}^C + \mathbf{F}_{ij}^D + \mathbf{F}_{ij}^R\right) \quad (1)$$

The conservative force of this expression are represented as (29)

$$\mathbf{F}_{ij}^C = A_{ij}\omega_c(r_{ij})\mathbf{e}_{ij} + B(\bar{\rho}_i + \bar{\rho}_j)\omega_d(r_{ij})\mathbf{e}_{ij} \quad (2)$$

where the first term in conservative force with a negative coefficient *A* stands for an attractive interaction within a range $r_c = 1$ as and the second term with *B* is the density-dependent repulsive interaction within a short range $r_d = 0.75$. There is a No-go theorem that constraints the condition for the parameters of the multi-component system, which means the force law is not conservative unless $B_{ij}$ is a constant matrix (32). Therefore, the repulsion parameters *B* are kept to 25 for all pairs in our simulation. The weight function is $\omega_c = 1 - r_{ij}/r_c$ for $r_{ij} \leq r_c$ and $\omega_c = 0$ for $r_{ij} > r_c$. Define the local density of repulsive term as $\rho_i = \sum_{i \neq j}\omega_\rho(r_{ij})$ for each particle, the generalized weight function is expressed as (29)

$$\omega_\rho(r_{ij}; r_d) = \frac{15}{2\pi r_d^3}\left(1 - r/r_d\right)^2 \quad (3)$$

The dissipative, and random forces are defined, respectively, as

$$\mathbf{F}_{ij}^{D} = -\gamma \omega_{D}(r_{ij})(\mathbf{e}_{ij} \cdot \mathbf{v}_{ij})\mathbf{e}_{ij} \tag{4}$$

$$\mathbf{F}_{ij}^{R} = \delta \omega_{R}(r_{ij})\xi_{ij}\Delta t^{-1/2}\mathbf{e}_{ij} \tag{5}$$

where $\gamma$ is the dissipative parameter and $\xi_{ij}$ is a random variable conforming to Gaussian distribution. The system satisfies the Gibbsian equilibrium and the fluctuation-dissipation theorem (33) if the dissipative parameter $\gamma$ and the amplitudes of random force $\delta$ satisfied $\delta^2 = 2\gamma k_B T$ and the weight function following $\omega_D(r) = [\omega_R(r)]^2$. The combined effect of dissipative force and random force acts as a thermostat.

### 2.1.2. CG model of phospholipid

In classic DPD coarse-grained strategy developed by Groot (26), one water bead (W) corresponds to 3 water molecules. Thus, the number $N_m = 3$ can be regarded as the CG degree of model. The CG model of DPPC has two hydrophilic head beads (H) representing the phosphate moiety and the choline moiety, two backbone beads (G) connecting the head and tail representing the glycerol linkage and four hydrophobic tail beads (T/C1) at each chain. Each tail bead corresponds to four carbon atoms (Fig. 1A). The only difference between the CG model of POPC and DPPC is the bead type of the second bead near the glycerol linkage on the main tail chain, due to the presence of unsaturated bond in POPC. Therefore, that carbon bead containing double bond was redefined as type T/C2. It was similar with the DPPC and POPC CG scheme adopted by Marrink in the MARTINI force field (23, 34).

Bonded interactions between connected beads are represented by a weak harmonic potential. Following Groot and Gao (26, 35), bonds are described by $V_{bond}(r) = \frac{1}{2} K_{bond}(r-r_0)^2$ with an equilibrium bond distance $r_0 = 0.65 r_c$ and a force constant of $K_{bond} = 200 k_B T / r_c^2$ is applied on all neighboring beads except that glycerin beads linkage ($r_0 = 0.55 r_c$). Angles are described by $V_{angle}(\theta) = \frac{1}{2} K_{angle}(\theta - \theta_0)^2$ for adjacent three beads. A force constant of $K_{angle} = 4 k_B T / rad^2$ and an equilibrium value of the angle $\theta_0 = 180°$ are applying on two head beads connected to one glycerin bead, second head bead connected to the first glycerin bead and tail bead, glycerin bead connected to two tail beads, and three consecutive tail beads; for two glycerin beads connected to head bead or tail bead, $K_{angle} = 4 k_B T / rad^2$ and $\theta_0 = 120°$; for the angles involving the cis double bond, $K_{angle} = 8 k_B T / rad^2$ and $\theta_0 = 120°$.

### 2.1.3. System setup

For calculations of the surface tension at the air-water interface, the simulation setup (Fig. 1B) included a small water cube 8×8×10 nm containing 4358 water beads in an 8×8×20 nm simulation box and a large water cube 30×30×10 nm containing 61290 water beads in a 30×30×30 nm simulation box. To determine the interaction parameters of the tail beads and water, 968 hexadecane (C4) molecules or 1291 dodecane (C3) were placed in an 8 × 8 × 20 box with an ensemble. In the surface tension simulation (Fig. 1B), an oil slab contained 728 hexadecane (C4) molecules and two boundary water slabs contained 4352 water beads were placed in a box of the same size.

For monolayer simulations, we simulated DPPC, POPC and mixed DPPC/POPC monolayers. The simulation setup of the monolayer included a water slab sandwiched by two air slabs and two symmetrical monolayers covered on two air-water interfaces (Fig. 1B). For mixed DPPC/POPC monolayers, DPPC and POPC were mixed with three molar ratios of 1:3, 1:1 and 3:1. For all the phospholipid monolayers above, the small system contained 444 molecules/monolayer and 21792 water beads; the large system contained 4000 molecules/monolayer and 196128 water beads. To obtain the surface pressure-area isotherms, a series of initial structures of lipid monolayers with different values of area per lipid (APL) were generated by controlling the lateral box size using PACKMOL package (36). The lateral size of the small box is in the range of 15−25 $r_c$ and the large box is in the range of 45−75 $r_c$.

### 2.1.4. Simulation details

All simulations were performed by the Meso-DPD module in LAMMPS package (37, 38). Note that the original Lucy kernel function in the MDPD package was replaced by equation (3) to calculate the local density. All simulations were performed in a three-dimensional box with periodic boundary conditions in all directions. The time step was set to be $\delta t = 0.01$ with MDPD time unit and the equilibration simulation was sustained by 1000000 timesteps for small systems and 10000000 timesteps for large systems to ensure the simulation convergence. The criterion for convergence is that the calculated surface tension remains constant over a long period of time. All simulations were carried out at 300 K using the NVT ensemble. The visualization of molecular configurations and simulation results was performed using VMD software (39).

*2.2. Constrained drop surfactometry experiments (CDS)*

### 2.2.1. Monolayer formation

DPPC (purity > 99%) and Chloroform ($CHCl_3$, purity > 99%) were purchased from Avanti Polar Lipids (Alabaster, AL). POPC (purity > 99%) was purchased from Sigma-Aldrich (St. Louis, MO). Phospholipids for the experiments were used without further purification. Mili-Q ultrapure water with pH = 7.4 from Millipore simplicity water purification system was used in all experiments. 20 mg of DPPC and POPC were dissolved in 20 mL of chloroform to form 1 mg/mL DPPC stock solution and POPC stock solution, respectively. DPPC/POPC mixture stock solutions were prepared according to the molecular weights of DPPC and POPC as 734.04 and 760.08. The molar ratios of DPPC:POPC are 1:3, 1:1, and 3:1. Subsequently, ultrasonic water bath was used to sonicate all the prepared stock solutions at 25°C and 40 kHz frequency for 5 minutes to obtain a series of uniformly mixed stock solutions. The droplet (~15 μL in volume) was constrained on a hydrophilic pedestal (~3 mm in diameter) that uses its knife-sharp edge to prevent film leakage and to maintain the droplet integrity. A small amount of stock solution (1 mg/mL) was spread onto the droplet by using a micro-syringe. Then the droplet could completely evaporate chloroform in one minute without interference.

### 2.2.2. Surface pressure−area isotherms measurement

The droplet was slowly expanded to increase the surface tension ($\gamma$) of the monolayer until it was close to the surface tension ($\gamma_0$) of pure water. The spread lipid monolayer was subsequently compressed at a rate of 0.005 cm$^2$/s and the real-time profile images of the droplet were directly displayed in the graphical user interface of the Axisymmetric Drop Shape Analysis (ADSA), processed and analyzed to generate a series of real-time surface tension measurement values to get the complete compression pressure-area isotherm (40-42). Each measurement was repeated three times to ensure the accuracy and repeatability of the experimental results. All measurements were carried out at $27 \pm 0.1\,°C$.

## 3. RESULTS AND DISCUSSION

### 3.1. The parameterization of water and oil (lipid tail)

We first calibrate the MDPD parameters for the water model by mapping the liquid phase properties of water to its actual physical properties and correlate the MDPD dimensionless units to the actual units. The interaction parameter $A$ of water beads is set to −50, following the previous MDPD water model that can reproduce the interface properties of water consistently (43). The simulation results in the equilibrium number density of water beads in the box is 6.8 $r_c^{-3}$, same in large box and small box. One water bead represents 3 water molecules and the volume of one water molecule is 30 $\text{Å}^3$, which means that one water bead in box occupies a real volume of 90 $\text{Å}^3$. According to the number density of the water beads in the box, the simulated characteristic length $r_c$ = 8.49 Å could be obtained, which determines the length scale of the system. In addition, the surface tension of water in MDPD unit is $12.4\,k_B T / r_c^2$. From the simple scaling relations, the calculated density and surface tension of water are 997 $\text{kg} \cdot \text{m}^{-3}$ and 71.2 mN/m at a room temperature of 300 K, which is consistent with the experimental results. After the characteristic length of the simulation is determined, the time scale of the simulation is obtained through mapping the calculated diffusion coefficient $D_{\text{bead}}$ of water to the experimental value $D_{\text{water}} = 2.43 \times 10^{-9}$ m$^2$ s$^{-1}$. The correspondence between the MDPD parameters in dimensionless units and the actual physical values is shown in Table 1.

**TABLE 1. Correspondence between the MDPD parameters in dimensionless units and the physical values.**

| MDPD | | MDPD → Real units | Physical units |
|---|---|---|---|
| Parameter | Value | | Value |
| Bead | 1 | $N_m$ | 3 H$_2$O |
| $r_c$ | 1 | $(\rho N_m V)^{1/3}$ | 8.49 Å |
| $\rho$ | 6.8 | $\rho N_m M / N_a r_c^3$ | 997 kg·m$^3$ |
| $\gamma$ | 12.4 | $\gamma k_B T / r_c^2$ | 71.2 mN·m$^{-1}$ |
| $p$ | 0.1 | $p k_B T / r_c^3$ | 6.75 MPa |
| $\delta t$ | 0.01 | $N_m D_{\text{bead}} r_c^2 / D_{\text{water}}$ | 0.32 ps |

Several important physical properties $r_c$, $\rho$, $\gamma$, $p$ and $\delta t$ correspond to the cutoff radius, density, surface tension, pressure, and time step, respectively. $V$ is the volume of one water molecule, $M$ is the molar weight of a water molecule, $N_a$ is Avogadro's number, $k_B$ is Boltzmann's constant, T is 300 K.

Given that the monolayer surface tension is mainly dominated by the interactions of the tail-air and head group-water, we calibrate the parameters of the tail and head beads dividually. The carbon atom in the saturated tail is similar with the one in the alkane compounds, such as hexadecane and dodecane. Thus, we compare the calculated bulk and surface properties of the alkanes with the experimental values to calibrate the parameters of the tail beads. The other beads of the lipid, including group beads, linkage beads and unsaturated beads, are calibrated by directly comparing the calculated and experimental pressure-area isotherms of the lipid monolayer, which is shown in the next section.

The hexadecane molecule model is divided into four CG beads and the dodecane is divided into three beads. The bond length between two adjacent carbon beads is set to $0.65\, r_c$, bonding stiffness is $200\, k_B T / r_c^2$, angular stiffness is $4\, k_B T / rad^2$ based on previous model (35). The attraction interaction parameter $A$ is appropriately adjusted by fitting the density of the oil and the surface tension of the oil-air and oil-water well with the experimental values.

When $A_{CC}$ is tuned to −28, the hexadecane/air surface tension is 26.3 mN/m and the density is 800 kg·m$^{-3}$; the dodecane-air surface tension is 23.6 mN/m and the density is 789 kg·m$^{-3}$. When $A_{WC}$ is set to −28.5, the hexadecane/water surface tension is 52.8 mN/m. These results are in good agreement with the experimental values (44). Surface tensions between different phases are shown as Table 2.

**TABLE 2. Interfacial Tension (mN/m) between different phases**

| System | MDPD | Martini | Experimental |
|---|---|---|---|
| Water-air | 71.2 | 45 | 72.0 |
| Dodecane-air | 23.6 | 25.3 | 24.0 |
| Hexadecane-air | 26.3 | 27.2 | 27.3 |
| Hexadecane-water | 52.8 | 55.2 | 53.0 |

*3.2. Surface pressure-area isotherms of the DPPC and POPC monolayers*

After determining the interaction parameters between carbon beads (T/C1) and water beads (W), we screen the parameters of other beads within a reasonable interval to determine the interaction parameters. The DPPC monolayer in the small system is simulated to obtain a group of pressure-area isotherms with a series of parameter sets that are compared with the experimental results (Fig. S1 A). In this way, we can determine a set of parameters that fits the calculated isotherm well with the experimental one. After parameters of the head beads and the linkage beads are calibrated, the parameterization of the T/C2 beads of POPC molecule can be obtained through the same methods (Fig. S1B). The non-bonded interaction parameters for all pairs of beads are shown in Table 3.

TABLE 3. MDPD beads non-bonded interaction parameters

| $A_{ij}$ | W | H | G | T/C1 | T/C2 |
|---|---|---|---|---|---|
| W | -50 | -48 | -41 | -28.5 | -33 |
| H | -48 | -36 | -31 | -23 | -23 |
| G | -41 | -31 | -31 | -31 | -27 |
| T/C1 | -28.5 | -23 | -31 | -28 | -26 |
| T/C2 | -33 | -23 | -27 | -26 | -33 |

The surface pressure-area isotherm is given by a series of surface pressure points corresponding to the relation: $\pi(a) = \gamma_0 - \gamma(a)$, where $\gamma_0$ denotes the surface tension of the air-water interface. The surface tension in the monolayer, $\gamma$, is calculated from the difference of the normal pressure $P_N$ and lateral pressure $P_L$ in the box according the Irving–Kirkwood (45) approach, expressed as $\gamma = (P_N - P_L) \cdot L_z / 2 = (P_{zz} - (P_{xx} + P_{yy})/2) \cdot L_z / 2$, where $L_z$ is the size of the box in the normal direction, $P_{xx}$, $P_{yy}$ and $P_{zz}$ are the ensemble-time average of pressure components in x, y and z directions. Each point of the surface pressure-area isotherm is obtained from one independent simulation at the constant area. The points of surface pressure-area isotherm are obtained when the calculated ensemble-time averaged surface tension stabilized, which means that the simulated phospholipid monolayer reaches a metastable or quasi-equilibrium state.

The calculated surface pressure-area isotherm of DPPC monolayer is shown in Fig. 2A with the isotherms determined by other techniques. There are four main phase regions in the calculated isotherm. When the area per lipid (APL) is larger than 0.9 nm$^2$, the monolayer in liquid-expanded (LE) phase coexists with pores (the corresponding snapshot is shown as Fig. 2B). At this stage, the surface pressure is reduced to near 0 mN/m and the APL is 0.9 nm$^2$, which is much more coincident with the experimental and AAMD results than the Martini force field. When the APL ranges from 0.9 to 0.65 nm$^2$, the DPPC monolayer is in LE phase with no pores and the lipid tails are mainly disordered (Fig. 2C). As the APL continues increasing (0.65-0.55 nm$^2$), the slope of the isotherm is distinctly changed and the DPPC monolayer is in the coexistence of LE and liquid condensed (LC) phase. In the AAMD and Martini simulations, the isotherm reaches to a plateau at the LE-LC phase and the slope of the isotherm fluctuates around zero. In our simulations, the slope of the isotherm from the LE to LC phase smoothly changes, which is closer to the experimental results. For the APL ranging from 0.55-0.45 nm$^2$, the monolayer is in the LC phase and the tails are ordered (Fig. 2D). As the surface pressure reaches to 70 mN/m, that is the surface tension is 0 mN/m, the monolayer is unstable and collapses with the lipids extruded to the water phase to form a micelle structure (Fig. 2E-F). In general, the pressure-area isotherm with the corresponding monolayer morphology determined by the MDPD simulations is more coincident with the experimental results than the Martini force field (12). In addition, the calculated pressure-area isotherms of different sizes of monolayers are consistent, which is different from the size effect existing in the traditional AAMD simulations of the lipid monolayer.

Slightly different from DPPC molecule, POPC molecule has one unsaturated C=C bonds in one of its tails and increases the disorder of the tails. MDPD simulated and CDS experimental pressure-area isotherms are shown in Fig. 3A and compared with several experimental isotherms (46-49) at the same temperature. At the range of the APL from 0.45−1.15 nm$^2$, the surface pressure of the POPC monolayer is overall higher than that of the DPPC monolayer at the same APL (Fig. S2). The POPC monolayer is more stable than DPPC monolayer at the low surface pressure and the pore formation starts at APL=1.0 nm$^2$ (Fig. 3B), which is similar with the previous experiments. Besides, POPC monolayer is always in the LE phase at 300 K (Fig. 3C and D), due to the unsaturation chains in the POPC molecules. As the lipid density becomes higher, the POPC monolayer collapses at APL= 0.42 nm$^2$ and the surface pressure of 61.8 mN/m reaching to a plateau in the isotherm that is in line with our CDS experiment.

*3.3. Mechanical properties of the DPPC and POPC monolayers*

From the slopes of the pressure-area isotherms, we can obtain the area compressibility modulus $Cs^{-1}$, which is the reciprocal of the compressibility $Cs$ and equivalent to elasticity. The compressibility can be expressed as:

$$Cs = -\frac{1}{A}\left(\frac{\partial A}{\partial \pi}\right)_T$$

where $A$ and $\pi$ are the APL and surface pressure, respectively. Typical area compressibility modulus of DPPC monolayers measured by experiments and simulations is in the range of 10−50 mN/m in the LE phase and 100−250 mN/m in the LC phase (50, 51). We summarized our simulated and experimental data, as well as the data in the existing literatures in Table 4. The area compressibility moduli of the LC and LE phase of the monolayer are approximated by linear regression from the pressure-area isotherm. The calculated compressibility moduli of the MDPD DPPC monolayer in the LC and LE phase are about 310 and 55 mN/m, which is close to the typical experimental results (300 mN/m in LC phase and 55 mN/m) and is much smaller than the Martini force field. POPC monolayer is mainly at the LE phase in the APL of 0.55-1.0 nm$^2$, and the slope of the isotherm obviously changes at such APL range. Thus, the calculated compressibility modulus of the POPC ranges from 30 mN/m to 85 mN/m related to the APL and the calculated values are similar with the CDS experiments as well as previous experiments. At the same APL, the modulus of POPC monolayer is much smaller than the one of the DPPC monolayer, which is in accordance with the experimental results.

**TABLE 4.** Experimental and simulated area compressibility moduli

|  | Temperature/ K | Phase | $C_s^{-1}$/mN/m | Area/ Å² |
|---|---|---|---|---|
| **DPPC Monolayers** | Varies |  |  | Varies |
| MDPD Large | 300 | LC | ~ 263 | ~ 49.4 |
| MDPD Small | 300 | LC | ~ 260 | ~ 49.4 |
| CDS | 300 | LC | ~ 254 | ~ 43.9 |
| Tieleman et al. (12) (Martini) | 300 | LC | ~ 786 | ~ 47.6 |
| Javanainen et al. (52) (AAMD) | 298 | LC | ~ 315 | ~ 50.2 |
| Crane et al. (15) (Exp) | 298 | LC | ~ 272 | ~ 45.0 |
| MDPD Large | 300 | LE | ~ 46 | ~ 70 |
| MDPD Small | 300 | LE | ~ 46 | ~ 70 |
| CDS | 300 | LE | ~ 43 | ~ 70.9 |
| Tieleman et al. (12) (Martini) | 300 | LE | ~ 243 | ~ 58.9 |
| Javanainen et al. (52) (AA) | 298 | LE | ~ 58 | ~ 78.2 |
| Crane et al. (15) (Exp) | 298 | LE | ~ 32 | ~ 79.7 |
| **POPC Monolayers** | Varies |  |  | Varies |
| MDPD Large | 300 | LE | ~ 152 | ~ 52.6 |
|  | 300 | LE | ~ 36 | ~ 80 |
| MDPD Small | 300 | LE | ~ 156 | ~ 50.8 |
|  | 300 | LE | ~ 43 | ~ 80 |
| CDS | 300 | LE | ~ 110 | ~ 52 |
|  | 300 | LE | ~ 58 | ~ 70 |
| Olżyńska et al. (46) (Exp) | 298 | LE | ~ 86 | ~ 60 |
|  | 298 | LE | ~ 38 | ~ 90 |
| Brown et al. (47) (Exp) | 297 | LE | ~ 114 | ~ 58 |
|  | 297 | LE | ~ 45 | ~ 90 |
| Volinsky et al. (48) (Exp) | 298 | LE | ~ 85 | ~ 55 |
|  | 298 | LE | ~ 47 | ~ 80 |
| Prenner et al. (49) (Exp) | 298 | LE | ~ 119 | ~ 55 |
|  | 298 | LE | ~ 47 | ~ 80 |

*3.4. Molecular structure of the DPPC and POPC monolayer*

During the expression and expansion, the lipid monolayer at the air-water interface exhibits different molecular structures. In this section, we examine the molecular structure of the DPPC and POPC monolayers at different stages. First, we show the molecular structure transformation of the DPPC and POPC monolayer at the rupture and collapse stages under the extremely high and low surface tension in Fig. 4.

For the DPPC monolayer, the poration of the monolayer starts at APL=0.9 nm$^2$ with the stable pores and these pores will become bigger with the increase of the APL to 1.0 nm$^2$. Pores are also observed in experiments at the gas phase (>0.9 nm$^2$) and in AAMD simulations at APL=1.0 nm$^2$, which are similar with our calculated values. Compared with DPPC monolayer, the POPC monolayer is more stable at the high surface tension and starts to form pores at APL=1.0 nm$^2$. This is because the work of formation of a round pore of radius r equals to $2\pi r \Gamma - \pi r^2 \sigma$ ($\Gamma$ is the line tension at the edge of the pore and $\sigma$ is the surface tension) (53) and the POPC molecule has a larger line tension due to the unsaturated bead. At the collapse stage, the DPPC monolayer collapse from the buckling and fold to form the bilayer structure in the water phase. As the protruded bilayer length increases, the bilayer becomes unstable and detaches from the monolayer forming a flat bilayer structure in the water phase, which is similar with the previous CGMD simulations (54). Moreover, our model suggests that the POPC monolayer can collapse at a lower surface pressure because of its lower elastic modulus.

Except for the rupture and collapse stages, although the lipid monolayer is always flat under the expression and expansion, the lipid monolayer exhibits different molecular structures at the different surface tensions. Order parameter of the lipid tails is an important property to quantitatively describe the orientation of the lipids and the phase separation. The order parameter is defined as (20)

$$S_z = \frac{3}{2}\langle \cos^2 \theta_n \rangle - \frac{1}{2}$$

where $\theta_n$ is the angle between the calculated molecular axis connecting the n−1 and n+1 sites of the hydrocarbon chain and the normal z-axis of the monolayer. At a low surface tension (APL=0.45, 0.47 nm$^2$), DPPC monolayer is in the LC phase and the corresponded order parameter is much larger than the one in the LE phase (APL ≥ 0.51 nm$^2$). These calculated values of the order parameter are quantitively in line with the AAMD simulations (46, 55). The POPC monolayer has a bond angle of 120° in the unsaturated tail chain thus leading an obvious decrease in the order parameter on the third bead, while the order parameter curve of DPPC is relatively smooth. Moreover, the order parameters of POPC monolayer are far smaller than that of the DPPC monolayer at the same APL. Even at the APL=0.47 nm$^2$, the order parameter of POPC is slightly smaller than that of DPPC at APL=0.51 nm$^2$, indicating that the POPC monolayer is always in the LE phase at the 300 K. This is also coincident with the molecular structure of the DPPC and POPC molecules (Fig. 5C and D). We also check the bead density distributions of the monolayer at the z-axis (Fig. S3), where the POPC monolayer shows a much smaller thickness than the DPPC monolayer at the same APL. This indicates that the POPC molecules are more likely to parallelly arranged at the air-water interface, which is in accordance with the order parameters.

*3.5. Pressure-area isotherms of mixed DPPC/POPC monolayers*

The natural lung surfactant monolayer has complex and unique biophysical properties due to their complex lipid components. Therefore, modeling towards the real multi-component lipid monolayer is crucial for understanding the biophysics of the lung surfactant monolayer. Here, we study the experimental and simulated isotherms of three mixed DPPC/POPC monolayers of different ratios in Fig. 6. The isotherms of the simulated and experimental mixed DPPC/POPC monolayers are between the isotherms of the pure DPPC and POPC monolayers. With the increase of the ratio of POPC, the isotherm of the mixed monolayer is closer to the pure POPC. Although the mixed monolayer has the intermediate behavior of two lipids, the shapes of the simulated isotherms of the mixed monolayers are mainly similar with the isotherm of pure POPC monolayer, which is as the same trend with the experimental results. However, it should be noted that at the APL > 0.5 nm$^2$, the calculated isotherms of mixed monolayer have a larger slope with a steep increase of the pressure, while the pressure in the experiment isotherms are slowed increased that leads to a slight deviation in the compressibility between the simulations and experiments. Generally, our model is capable to capture the features of the mixed lipid monolayer and the parameterization scheme could be applied to more types of phospholipids, which can be used for mimicking the mesoscopic behavior of natural lipid monolayers.

It should be noted that although our MDPD model is capable to nearly quantitively simulate the phospholipid monolayers at the mesoscale, there are still some limitations that should be noted. First, for simulating a large system of DPPC monolayers with a horizontal size of 63 nm × 63 nm and a total bead number of 346000, it takes about 25 hours to run 10,000,000 steps on 224 CPU cores in 8 nodes (3200 ns). Therefore, our simulation scale is limited to hundreds of nanometers and microseconds. Second, our MDPD model does not include any electrostatic interactions. Third, the bond and angle parameters are referred to previous DPD lipid models, which is worth optimizing for more accurate monolayer simulations.

## 4. CONCLUSION

Although phospholipid monolayers at the air-water interface have been widely studied using experimental and computational methods, it is still difficult to quantitively model the physical properties of the phospholipid monolayers at the mesoscopic scale. Here, we present a MDPD model for two commonly used phospholipids, DPPC and POPC. Using the parameterization scheme that based on mapping the calculated physical properties to the experimental values, we reproduce the pressure-area isotherms of the phospholipid monolayers in MDPD simulations with a good fitting with our CDS experiments. Moreover, the mechanical property and molecular structures of phospholipid monolayers are quantitively and qualitatively in line with the experiments and AAMD simulations. Compared with the experimental results, this model can also capture the sensitive changes in the pressure isotherms of the mixed DPPC/POPC monolayers by altering the mixed ratios of the components. These results demonstrate our model can be applied for mesoscale phospholipid simulations at the air-water interface.

**FIGURES AND CAPTIONS**

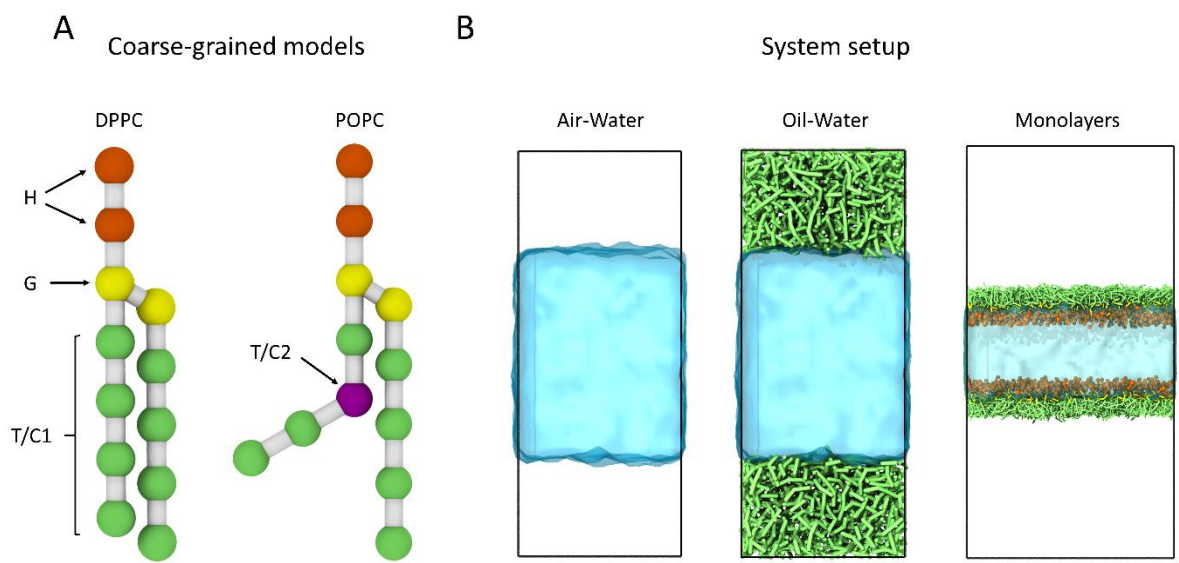

**FIGURE 1.** (A) CG MDPD models for DPPC and POPC molecules. Orange, yellow, and green beads stand for head (H), glycerol (G) and tail (T) beads. (B) Setups for calculations of water-air, water-oil, and phospholipid monolayer surface tensions.

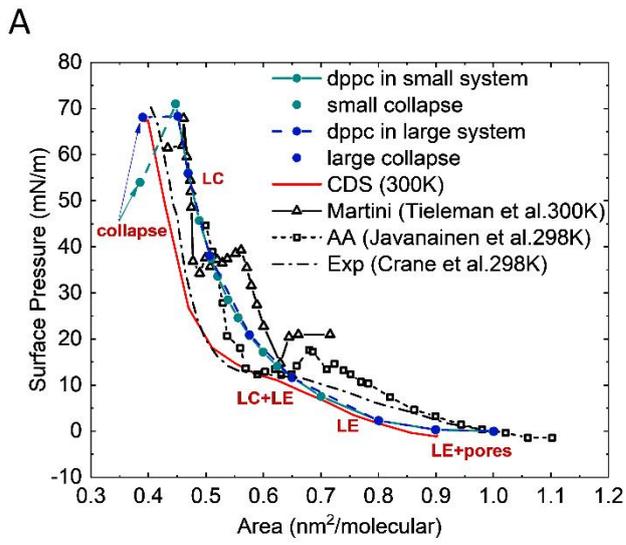
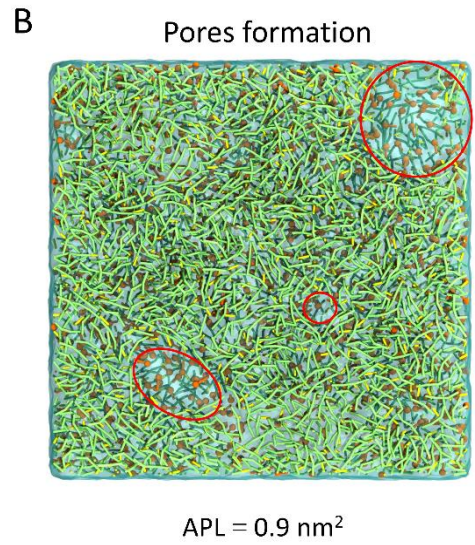

A. Surface pressure vs Area isotherm with labels: collapse, LC, LC+LE, LE, LE+pores. Legend: dppc in small system, small collapse, dppc in large system, large collapse, CDS (300K), Martini (Tieleman et al.300K), AA (Javanainen et al.298K), Exp (Crane et al.298K).

B. Pores formation. APL = 0.9 nm²

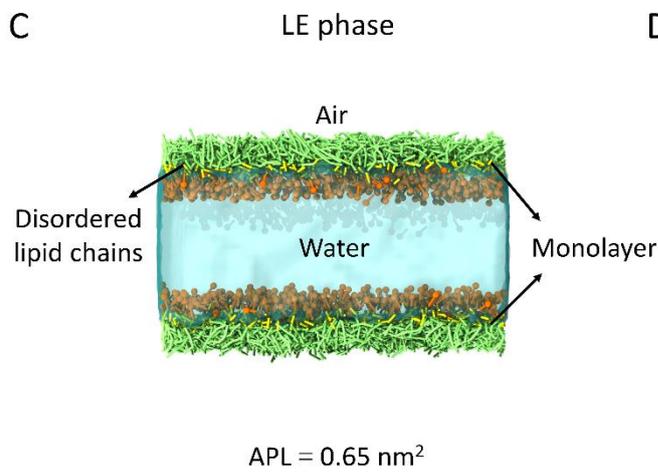
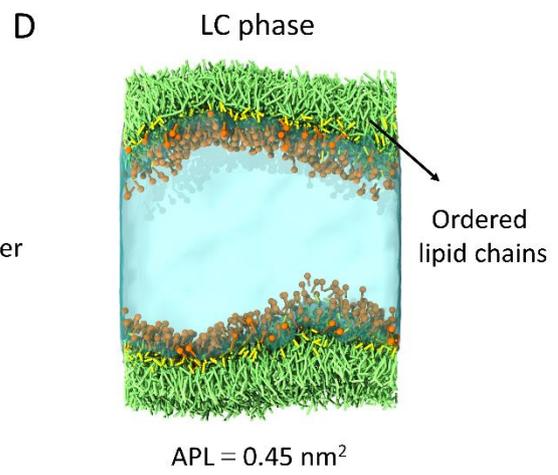

C. LE phase — Disordered lipid chains, Air, Water, Monolayer. APL = 0.65 nm²

D. LC phase — Ordered lipid chains. APL = 0.45 nm²

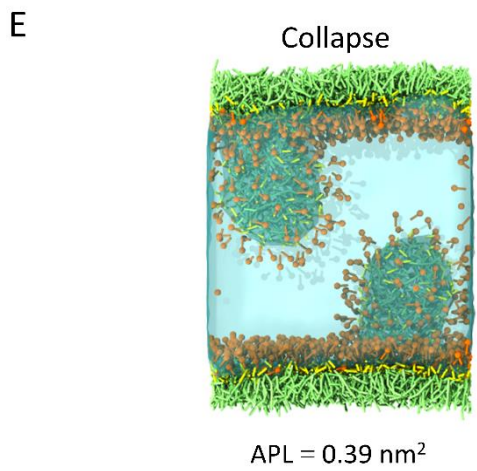
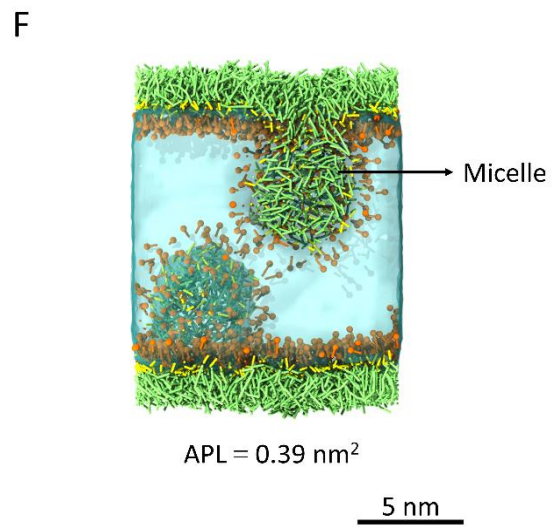

E. Collapse. APL = 0.39 nm²

F. Micelle. APL = 0.39 nm²

5 nm

**FIGURE 2**. (A) Surface pressure–area isotherms of pure DPPC monolayers from MDPD simulations and CDS experiments, compared with existing literature data including all-atom simulations, Martini simulations and experiments (12, 15, 52). Note that the isotherm in Martini simulations is corrected by shifting along the y axis. (B) The top view of the hole formation in the DPPC monolayer in the small system. (C) Side view of DPPC monolayers with disordered tail lipid chains in non-porous LE phase. (D) Side view of DPPC monolayers with ordered tail lipid chains in LC phase with ripples appeared. (E) Side view of the DPPC monolayer collapsed to form micelles in the water phase. (F) Sectional view of one of the DPPC micelles. The corresponding APLs are given below the snapshots.

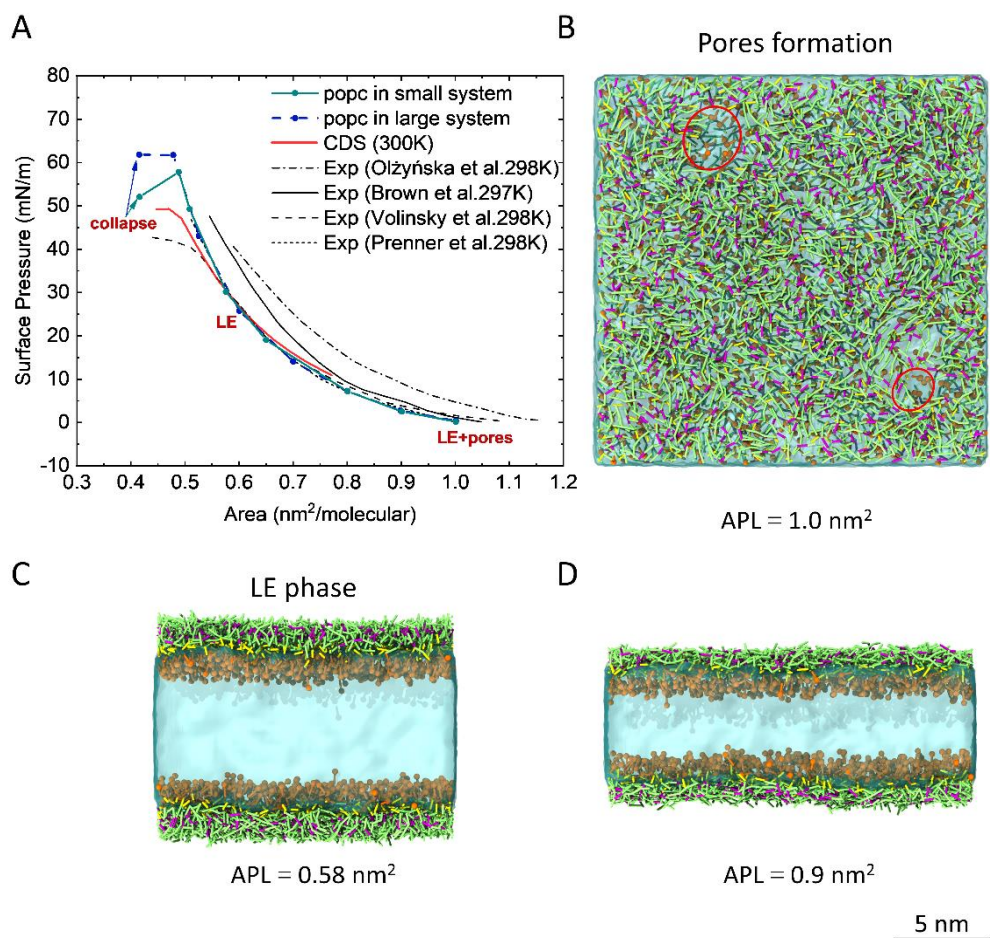

**FIGURE 3**. (A) Surface pressure–area isotherms of pure POPC monolayers from MDPD simulations, and CDS experiments, and existing experiments. (B) The top view of the hole formation in the POPC monolayer. (C) and (D) are the side views of POPC monolayers with disordered lipid tail chains in different areas of the LE phase.

A

DPPC monolayer

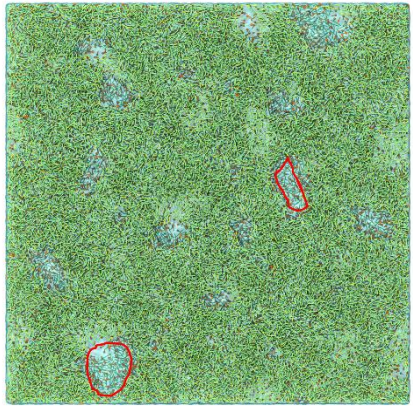 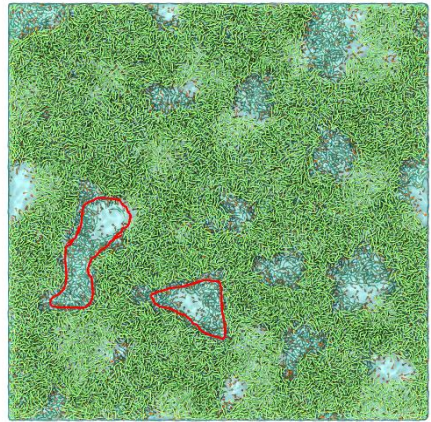

APL = 0.9 nm$^2$  APL = 1.0 nm$^2$

B

POPC monolayer

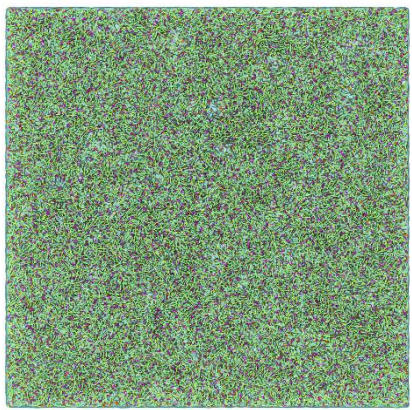 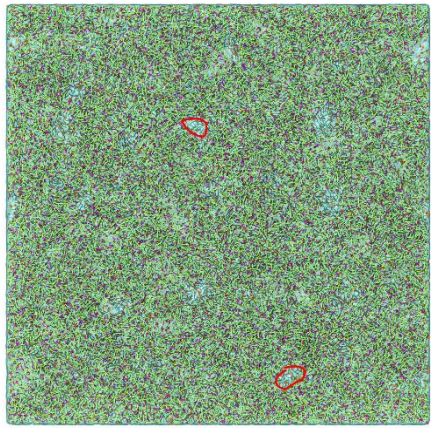

APL = 0.9 nm$^2$  APL = 1.0 nm$^2$

10 nm

C

Forming bilayer folds    Detaching from the monolayer

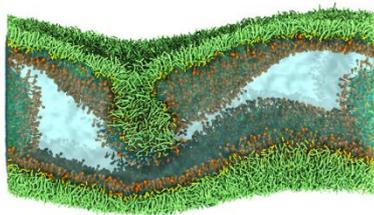 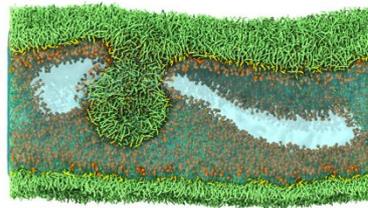

3 ns    19 ns

Bilayer formation    Stable flat circular bilayers

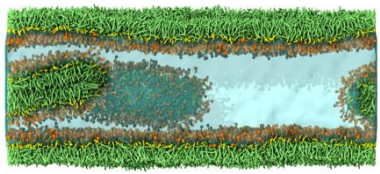 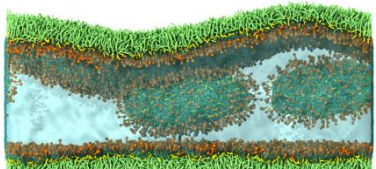

323 ns    1011 ns

APL = 0.39 nm$^2$    10 nm

**FIGURE 4**. Snapshots for the simulations of DPPC monolayers in the large system at the collapse and rupture stages. (A) and (B) Pore formation of DPPC and POPC monolayer at APL=0.9 and 1.0 nm$^2$ in the top view. (C) Structure transformation of the DPPC monolayer at the collapse stage. Forming bilayer folds (*top left*); Detaching from the monolayer (*top right*); The monolayers collapse forming two flat circular bilayers in the water phase (*bottom right*); Sectional view of the structure of one of the extruded DPPC bilayers (*bottom left*).

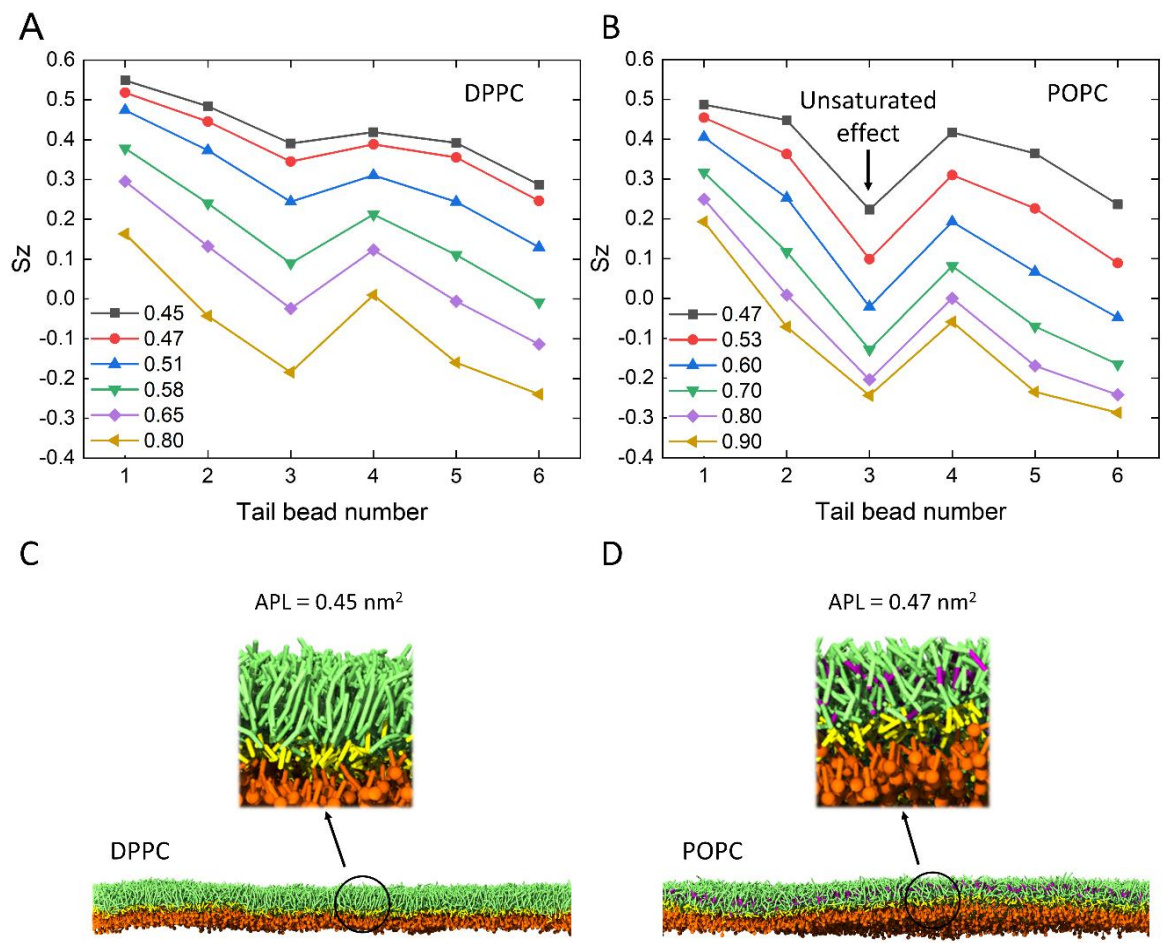

**FIGURE 5**. (A) and (B) Order parameters of tail beads in the DPPC and POPC monolayers at various APLs. The bead number stands for the bead serial in the tails, where 1-3 stand for the first three beads in the left tail and 4-6 stand for the first three beads in the right tail. The last bead in each tail is not included in order parameter calculations. (C) and (D) The molecular structure of the DPPC and POPC molecules at a similar APL.

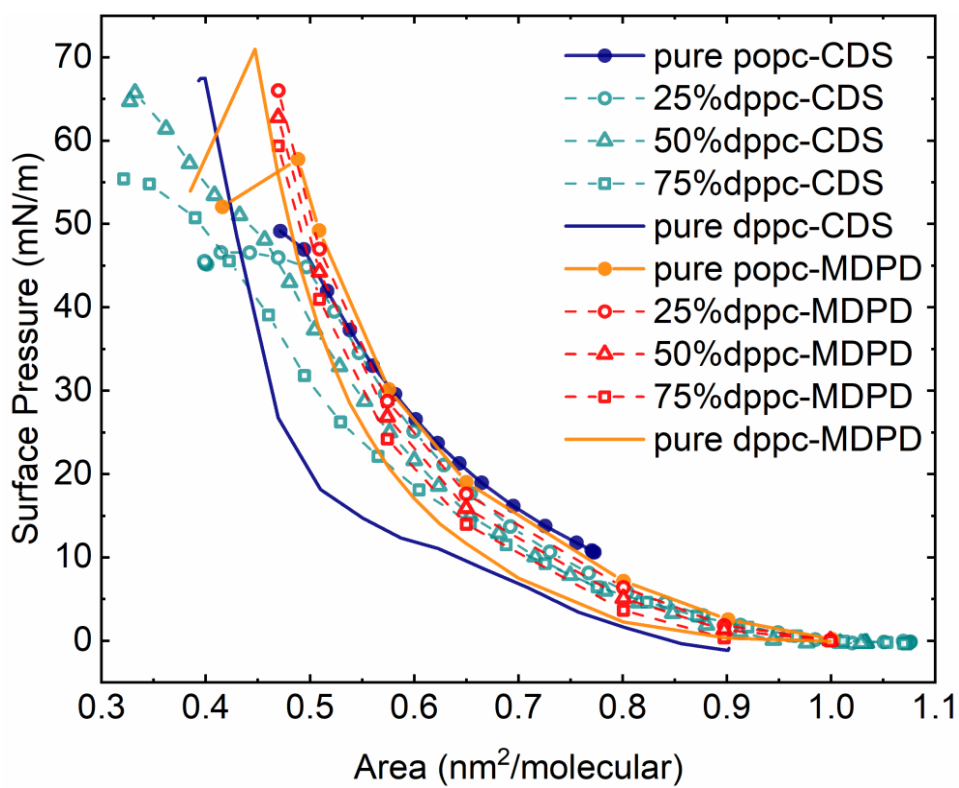

**FIGURE 6**. Surface pressure–area isotherms of three mixed DPPC/POPC monolayers and pure (DPPC, POPC) monolayers. The mixed ratio of DPPC molecules in the mixed monolayers are 25%, 50%, and 75%. The orange and blue solid lines represent the isotherms calculated and experimental DPPC monolayers. The orange and blue solid lines with solid circles represent the isotherms calculated and experimental of POPC monolayers. The red and deep cyan dot lines with hollow circles, triangles, and cubes represent the calculated and experimental isotherms of the mixed monolayer with the 25%, 50%, and 75% DPPC.

# AUTHOR CONTRIBUTIONS

X. B. and G. H. designed the research. Y. Z. performed the simulations and experiments. Y. Z. and X. B. analyzed and interpreted the data. Y. Z., X. B., and G. H. wrote the manuscript.

# ACKNOWLEDGEMENTS

This work was supported financially by the Natural Science Foundation of China (Grant Nos. 11832017, 11572334, 11772343), the Chinese Academy of Sciences Key Research Program of Frontier Sciences (QYZDB-SSW-JSC036), and the Chinese Academy of Sciences Strategic Priority Research Program (XDB22040403). The MDPD simulations were performed on TianHe-1(A) at the National Supercomputing Center in Tianjin.

# REFERENCES

1. Gruen, D. W. R., and J. Wolfe. 1982. Lateral Tensions and Pressures in Membranes and Lipid Monolayers. Biochimica Et Biophysica Acta 688(2):572-580.

2. Jahnig, F. 1984. Lipid Exchange between Membranes. Biophysical Journal 46(6):687-694.

3. Nagle, J. F. 1986. Theory of Lipid Monolayer and Bilayer Chain-Melting Phase-Transitions. Faraday Discuss 81:151-162.

4. Feng, S. S. 1999. Interpretation of mechanochemical properties of lipid bilayer vesicles from the equation of state or pressure-area measurement of the monolayer at the air-water or oil-water interface. Langmuir 15(4):998-1010.

5. Goerke, J. 1974. Lung surfactant. Biochimica et Biophysica Acta (BBA)-Reviews on Biomembranes 344(3-4):241-261.

6. Marsh, D. 1996. Lateral pressure in membranes. Bba-Rev Biomembranes 1286(3):183-223.

7. Schindler, H. 1979. Exchange and Interactions between Lipid Layers at the Surface of a Liposome Solution. Biochimica Et Biophysica Acta 555(2):316-336.

8. Walters, R. W., R. R. Jenq, and S. B. Hall. 2000. Distinct steps in the adsorption of pulmonary surfactant to an air-liquid interface. Biophysical Journal 78(1):257-266.

9. Bai, X., L. Xu, J. Y. Tang, Y. Y. Zuo, and G. Hu. 2019. Adsorption of Phospholipids at the Air-Water Surface. Biophys J 117(7):1224-1233.

10. Alonso, C., T. Alig, J. Yoon, F. Bringezu, H. Warriner, and J. A. Zasadzinski. 2004. More than a monolayer: Relating lung surfactant structure and mechanics to composition. Biophysical Journal 87(6):4188-4202.

11. Parra, E., and J. Perez-Gil. 2015. Composition, structure and mechanical properties define performance of pulmonary surfactant membranes and films. Chemistry and Physics of Lipids 185:153-175.

12. Baoukina, S., L. Monticelli, S. J. Marrink, and D. P. Tieleman. 2007. Pressure-area isotherm of a lipid monolayer from molecular dynamics simulations. Langmuir 23(25):12617-12623.

13. Zhang, H., Q. H. Fan, Y. E. Wang, C. R. Neal, and Y. Y. Zuo. 2011. Comparative study of clinical pulmonary surfactants using atomic force microscopy. Bba-Biomembranes 1808(7):1832-1842.

14. Langmuir, I. 1916. The constitution and fundamental properties of solids and liquids Part I Solids. Journal of the American Chemical Society 38:2221-2295.

15. Crane, J. M., G. Putz, and S. B. Hall. 1999. Persistence of phase coexistence in disaturated phosphatidylcholine monolayers at high surface pressures. Biophysical Journal 77(6):3134-3143.

16. Ahuja, R. C., and D. Mobius. 1992. Photophysical Properties of a Pyrene-Labeled Phospholipid in Matrix Monolayers at the Gas Water Interface. Langmuir 8(4):1136-1144.

17. Baldyga, D. D., and R. A. Dluhy. 1998. On the use of deuterated phospholipids for infrared spectroscopic studies of monomolecular films: a thermodynamic analysis of single and binary component phospholipid monolayers. Chemistry & Physics of Lipids 96(1-2):81-97.

18. Duncan, S. L., I. S. Dalal, and R. G. Larson. 2011. Molecular dynamics simulation of phase transitions in model lung surfactant monolayers. Biochim Biophys Acta 1808(10):2450-2465.

19. Zhang, S. Y., and X. B. Lin. 2019. Lipid Acyl Chain cis Double Bond Position Modulates Membrane Domain Registration/Anti-Registration. Journal of the American Chemical Society 141(40):15884-15890.

20. Baoukina, S., E. Mendez-Villuendas, and D. P. Tieleman. 2012. Molecular View of Phase Coexistence in Lipid Monolayers. Journal of the American Chemical Society 134(42):17543-17553.

21. Kaznessis, Y. N., S. T. Kim, and R. G. Larson. 2002. Simulations of zwitterionic and anionic phospholipid monolayers. Biophysical Journal 82(4):1731-1742.

22. Vanommeslaeghe, K., E. Hatcher, C. Acharya, S. Kundu, S. Zhong, J. Shim, E. Darian, O. Guvench, P. Lopes, and I. Vorobyov. 2010. CHARMM general force field: A force field for drug‐like molecules compatible with the CHARMM all‐atom additive biological force fields. J Comput Chem 31(4):671-690.

23. Marrink, S. J., H. J. Risselada, S. Yefimov, D. P. Tieleman, and A. H. de Vries. 2007. The MARTINI force field: coarse grained model for biomolecular simulations. J Phys Chem B 111(27):7812-7824.

24. Marrink, S. J., and D. P. Tieleman. 2013. Perspective on the Martini model. Chemical Society Reviews 42(16):6801-6822.

25. Hoogerbrugge, P. J., and J. M. V. A. Koelman. 1992. Simulating Microscopic Hydrodynamic Phenomena with Dissipative Particle Dynamics. Europhys Lett 19(3):155-160.

26. Groot, R. D., and K. L. Rabone. 2001. Mesoscopic simulation of cell membrane damage, morphology change and rupture by nonionic surfactants. Biophys J 81(2):725-736.

27. Espanol, P., and P. B. Warren. 2017. Perspective: Dissipative particle dynamics. J Chem Phys 146(15):150901.

28. Trofimov, S. Y., E. L. F. Nies, and M. A. J. Michels. 2002. Thermodynamic consistency in dissipative particle dynamics simulations of strongly nonideal liquids and liquid mixtures. Journal of Chemical Physics 117(20):9383-9394.

29. Warren, P. B. 2003. Vapor-liquid coexistence in many-body dissipative particle dynamics. Phys Rev E Stat Nonlin Soft Matter Phys 68(6 Pt 2):066702.

30. Zhou, P., J. Hou, Y. Yan, J. Wang, and W. Chen. 2019. Effect of Aggregation and Adsorption Behavior on the Flow Resistance of Surfactant Fluid on Smooth and Rough

Surfaces: A Many-Body Dissipative Particle Dynamics Study. Langmuir 35(24):8110-8120.

31. Ghoufi, A., J. Emile, and P. Malfreyt. 2013. Recent advances in Many Body Dissipative Particles Dynamics simulations of liquid-vapor interfaces. Eur Phys J E Soft Matter 36(1):10.

32. Warren, P. B. 2013. No-go theorem in many-body dissipative particle dynamics. Phys Rev E Stat Nonlin Soft Matter Phys 87(4):045303.

33. Espanol, P., and P. Warren. 1995. Statistical-Mechanics of Dissipative Particle Dynamics. Europhys Lett 30(4):191-196.

34. Marrink, S. J., A. H. de Vries, and A. E. Mark. 2004. Coarse Grained Model for Semiquantitative Lipid Simulations. The Journal of Physical Chemistry B 108(2):750-760.

35. Li, Y., H. Yuan, A. von dem Bussche, M. Creighton, R. H. Hurt, A. B. Kane, and H. Gao. 2013. Graphene microsheets enter cells through spontaneous membrane penetration at edge asperities and corner sites. Proceedings of the National Academy of Sciences of the United States of America 110(30):12295-12300.

36. Martinez, L., R. Andrade, E. G. Birgin, and J. M. Martinez. 2009. PACKMOL: A Package for Building Initial Configurations for Molecular Dynamics Simulations. J Comput Chem 30(13):2157-2164.

37. Li, Z., G. H. Hu, Z. L. Wang, Y. B. Ma, and Z. W. Zhou. 2013. Three dimensional flow structures in a moving droplet on substrate: A dissipative particle dynamics study. Physics of Fluids 25(7).

38. Plimpton, S., P. Crozier, and A. Thompson. 2007. LAMMPS-large-scale atomic/molecular massively parallel simulator. Sandia National Laboratories 18:43.

39. Humphrey, W., A. Dalke, and K. Schulten. 1996. VMD: Visual molecular dynamics. J Mol Graph Model 14(1):33-38.

40. Valle, R. P., T. Wu, and Y. Y. Zuo. 2015. Biophysical influence of airborne carbon nanomaterials on natural pulmonary surfactant. Acs Nano 9(5):5413-5421.

41. Yu, K., J. Yang, and Y. Y. Zuo. 2016. Automated Droplet Manipulation Using Closed-Loop Axisymmetric Drop Shape Analysis. Langmuir 32(19):4820-4826.

42. Yang, J. L., K. L. Yu, and Y. Y. Zuo. 2017. Accuracy of Axisymmetric Drop Shape Analysis in Determining Surface and Interfacial Tensions. Langmuir 33(36):8914-8923.

43. Ghoufi, A., and P. Malfreyt. 2011. Mesoscale modeling of the water liquid-vapor interface: a surface tension calculation. Phys Rev E Stat Nonlin Soft Matter Phys 83(5 Pt 1):051601.

44. Queimada, A. J., I. M. Marrucho, J. A. P. Coutinho, and E. H. Stenby. 2005. Viscosity and Liquid Density of Asymmetric n-Alkane Mixtures: Measurement and Modeling. International Journal of Thermophysics 26(1):47-61.

45. Irving, J., and J. G. Kirkwood. 1950. The statistical mechanical theory of transport processes. IV. The equations of hydrodynamics. The Journal of chemical physics 18(6):817-829.

46. Olzynska, A., M. Zubek, M. Roeselova, J. Korchowiec, and L. Cwiklik. 2016. Mixed DPPC/POPC Monolayers: All-atom Molecular Dynamics Simulations and Langmuir Monolayer Experiments. Biochim Biophys Acta 1858(12):3120-3130.

47. Brown, R. E., and H. L. Brockman. 2007. Using monomolecular films to characterize lipid lateral interactions. Lipid Rafts. Springer, pp. 41-58.

48. Volinsky, R., R. Paananen, and P. K. Kinnunen. 2012. Oxidized phosphatidylcholines promote phase separation of cholesterol-sphingomyelin domains. Biophys J 103(2):247-254.

49. Prenner, E., G. Honsek, D. Honig, D. Mobius, and K. Lohner. 2007. Imaging of the domain organization in sphingomyelin and phosphatidylcholine monolayers. Chem Phys Lipids 145(2):106-118.

50. Kodama, M., O. Shibata, S. Nakamura, S. Lee, and G. Sugihara. 2004. A monolayer study on three binary mixed systems of dipalmitoyl phosphatidyl choline with cholesterol, cholestanol and stigmasterol. Colloid Surface B 33(3-4):211-226.

51. Duncan, S. L., and R. G. Larson. 2008. Comparing experimental and simulated pressure-area isotherms for DPPC. Biophys J 94(8):2965-2986.

52. Javanainen, M., A. Lamberg, L. Cwiklik, I. Vattulainen, and O. H. S. Ollila. 2018. Atomistic Model for Nearly Quantitative Simulations of Langmuir Monolayers. Langmuir 34(7):2565-2572.

53. Burgess, S., A. Vishnyakov, C. Tsovko, and A. V. Neimark. 2018. Nanoparticle-Engendered Rupture of Lipid Membranes. Journal of Physical Chemistry Letters 9(17):4872-4877.

54. Shinoda, W., R. DeVane, and M. L. Klein. 2010. Zwitterionic Lipid Assemblies: Molecular Dynamics Studies of Monolayers, Bilayers, and Vesicles Using a New Coarse Grain Force Field. Journal of Physical Chemistry B 114(20):6836-6849.

55. Rose, D., J. Rendell, D. Lee, K. Nag, and V. Booth. 2008. Molecular dynamics simulations of lung surfactant lipid monolayers. Biophys Chem 138(3):67-77.